\documentclass[prl,
                twocolumn,
                preprintnumbers,
                amsmath,
                amssymb,
                showpacs,
                superscriptaddress]{revtex4-1}
\usepackage{graphicx} 
\usepackage{dcolumn}  
\usepackage{bm}       
\usepackage{amsfonts}
\usepackage{dsfont}
\usepackage{mathtools}
\usepackage[colorlinks=true,linkcolor=blue,citecolor=red,urlcolor=blue]{hyperref}
\usepackage{color}
\newcommand{\eq}[1]{Eq.~(\ref{#1})}
\newcommand{\ful}{\mbox{C$_{\mbox{\scriptsize{60}}}$}}

\begin{document}

\title{Resonant Auger-intercoulombic hybridized decay in the photoionization of endohedral fullerenes}

\author{Mohammad H. Javani}
\affiliation{%
Department of Physics and Astronomy, Georgia State University, Atlanta, Georgia 30303, USA}

\author{Jacob B. Wise}
\affiliation{%
Department of Natural Sciences, Center for Innovation and Entrepreneurship, Northwest Missouri State University,
Maryville, Missouri 64468, USA}

\author{Ruma De}
\affiliation{%
Department of Natural Sciences, Center for Innovation and Entrepreneurship, Northwest Missouri State University,
Maryville, Missouri 64468, USA}

\author{Mohamed E. Madjet}
\affiliation{%
Qatar Environment and Energy Research Institute (QEERI), P.O. Box 5825, Doha, Qatar}

\author{Steven T. Manson}
\affiliation{%
Department of Physics and Astronomy, Georgia State University, Atlanta, Georgia 30303, USA}

\author{Himadri S. Chakraborty}
\email[]{himadri@nwmissouri.edu}
\affiliation{%
Department of Natural Sciences, Center for Innovation and Entrepreneurship, Northwest Missouri State University,
Maryville, Missouri 64468, USA}

\date{\today}

\pacs{61.48.-c, 33.80.Eh, 36.40.Cg}


\begin{abstract}
Considering the photoionization of Ar@$\ful$, we predict resonant femtosecond decays of both Ar and $\ful$ vacancies through 
the continua of atom-fullerene hybrid final states. The resulting resonances emerge from the interference between simultaneous 
autoionizing and intercoulombic decay (ICD) processes. For Ar $3s\rightarrow np$ excitations, these resonances are far stronger 
than the Ar-to-$\ful$ resonant ICDs, while for $\ful$ excitations they are strikingly larger than the corresponding Auger features. 
The results indicate the power of hybridization to enhance decay rates, and modify lifetimes and line profiles.
\end{abstract}

\maketitle 

Intercoulombic decay (ICD), originally predicted by Cederbaum {\em et al.}\ \cite{cederbaum1997firstTh} and observed initially for 
Ne clusters \cite{marburger2003firstExp}, is a unique, naturally abundant nonradiative relaxation pathway of a vacancy in atom A in a cluster or molecule. 
An outer electron of A fills the vacancy and the released energy, instead of emitting 
a second electron of A as in standard Auger ionization, transfers to a neighboring atom B {\em via} Coulomb interactions 
to ionize B. Repulsion between holes in A and B may lead to fragmentation. Over the last decade and a half, a wealth of theoretical 
\cite{averbukh2011revTh} and experimental \cite{hergenhahn2011revExp} research has gone into studying ICD processes
in weakly bound atomic systems. These involve 
the observation of ICD in rare gas dimers \cite{jahnke2004rareDimer}, rare gas clusters \cite{oehrwall2004rareClust}, surfaces \cite{grieves2011surface}, 
and small water droplets \cite{jahnke2010water1,mucke2010water2}. ICD followed by resonant Auger decay has been identified
in Ar dimers using momentum resolved electron-ion-ion coincidence spectroscopy \cite{okeeffe2013arDimer,kimura2013arDimer}. Ultrafast ICDs of a
dicationic monomer in a cluster to produce a cluster tricataion \cite{santra2003triCataion} or multiply excited homoatomic 
cluster \cite{kuleff2010multiExcited} were predicted. Also, time domain measurements of ICD in He \cite{trinter2013time1} and Ne 
\cite{schnorr2013time2} dimers have recently been achieved. Besides fundamental science contexts, low energy ICD electrons find potential 
medical applications in the treatment of malignant cells \cite{gokhberg2013MedApp}.

Of particular interest is the resonant ICD (RICD) where the precursor excitation to form an inner-shell vacancy is accomplished by 
promoting an inner electron to an excited state by an external stimulant, generally electromagnetic radiation \cite{barth2005ricdExp1,aoto2006ricdExp2}
or, more recently, charge-particle impact \cite{kim2013DimExp}. A theoretical study of RICD followed by Ne $2s\rightarrow np$ 
excitations in MgNe clusters  
suggested the leading contribution of RICD among other interatomic decay modes \cite{gokhberg2006ricdTh}. Photoelectron 
spectroscopy with Ne clusters for $2s\rightarrow np$ excitations measured the signature of RICD processes \cite{barth2005ricdExp1}. 
Similar excitations in the double photoionization of Ne dimers were utilized to observe RICD by tracking the formation of energetic 
Ne$^+$ fragments \cite{aoto2006ricdExp2}. Most recently, strong enhancement of the HeNe$^+$ yield, as He resonantly couples with the radiation,
is detected \cite{trinter13HeNe}, confirming an earlier prediction \cite{najjari10TwoCenter}.

Atoms confined in fullerene shells forming endofullerene compounds are particularly attractive natural laboratories
to study RICD processes. There are two compelling reasons for this: (i) such materials are highly stable, 
have low-cost sustenance at the room temperature and are enjoying a rapid improvement in synthesis techniques \cite{popov2013endoSynth}; 
and (ii) the effect of correlation of the central atom with the cage electrons have been predicted to spectacularly affect
the atomic photoionization \cite{madjet2007giant}. A first attempt to predict ICD in endofullerenes 
was made by calculating ICD rates for Ne@$\ful$ \cite{averbukh2006endo-icdTh}. While some speculation on the
role of Coulomb interaction mediated energy transfer from atom to fullerene to broaden Auger lines has been made \cite{korol2011,amusia2006},
no studies, theoretical or experimental, of RICD resonances in the ionization cross section of endofullerenes have been 
performed. Furthermore, ICD of endofullerene molecules can uncover effects not yet
known. This is because: (i) endofullerenes being spherical analogues of asymmetric dimers consisting of
an atom and a cluster can also induce reverse RICD processes, the decay of cluster innershell excitations through the continuum of the confined atom, 
of uniquely different character than the forward RICD; and (ii) possibilities of atom-fullerene hybridized final states, predicted
to exist abundantly in these systems \cite{chakraborty2009xe-hybrid,maser2012zn-hybrid}, can significantly alter the 
properties of intercoulombic processes.
\begin{figure}[h!]
\includegraphics[width=7.5cm]{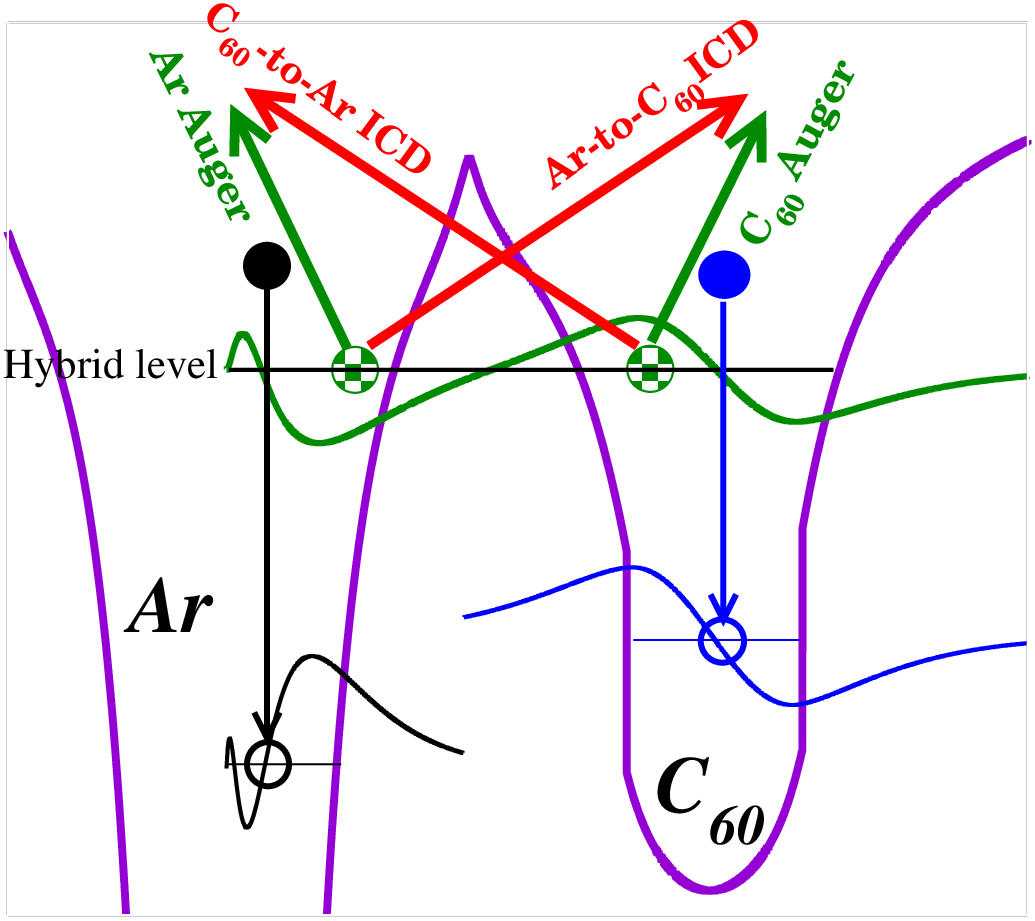}
	\caption{(Color online) Schematic of coherent mixings of one-center Auger decays (green) of core holes with corresponding ICDs (red)
	in the spectra of Ar-$\ful$ hybrid electrons.}
	\label{figure1}
\end{figure}

In this letter, we show that for an Ar atom endohedrally sequestered in $\ful$, ICD pathways of photo-generated innershell holes, both in the 
central atom and the fullerene, can coherently mix with degenerate {\em intracoulombic} Auger pathways to produce final states with 
$shared$ holes in atom-fullerene hybrid levels. Fig.\,1 presents a schematic of the process which illustrates this hitherto undetected mode 
that can be called the resonant hybrid Auger-intercoulombic decay or RHA-ICD.

A jellium based time-dependent local density approximation (TDLDA), with the Leeuwen and Baerends (LB) exchange-correlation functional  
to produce accurate asymptotic behavior \cite{van1994exchange} for ground and continuum states, is employed to
calculate the dynamical response of the system to the external electromagnetic field produced by the incident photon. 
The Ar nucleus is placed at the center of the sphere where the chemically inert noble gas atoms is known to localize. In solving the Kohn-Sham
equations to obtain the ground state wave function, a few essential optimizations were adopted \cite{madjet2010xeFull}. The model has enjoyed
earlier success in co-discovering with experimentalists a high energy plasmon resonance \cite{scully2005volumeExp}, interpreting the 
energy-dependent oscillations in $\ful$ valence photo-intensity data \cite{ruedel2002oscExp}, and predicting giant enhancements in the 
confined atom's photoresponse from the coupling with $\ful$ plasmons \cite{madjet2007giant}. Significant ground state hybridization of Ar $3p$
is found to occur with the $\ful$ $3p$ orbital, resulting in two levels, (Ar+$\ful$)$3p$ and  (Ar-$\ful$)$3p$, from
respectively, the symmetric and antisymmetric modes of mixing, the spherical analogs of bonding and 
antibonding states in molecules or dimers:
\begin{equation}\label{bound-hyb}
(\mbox{Ar}\!\pm\!\ful)3p = |\phi_\pm\rangle = \sqrt{\alpha}|\phi_{3p \scriptsize{\mbox{Ar}}}\rangle \!\pm\! \sqrt{1-\alpha}|\phi_{3p \scriptsize{\mbox{C}_{60}}}\rangle,
\end{equation} 
where $\alpha\simeq 0.5$. Such atom-fullerene hybridization was predicted earlier \cite{chakraborty2009hybrid} and detected 
in a photoemission experiment on multilayers of Ar@$\ful$ \cite{morscher2010strong}. In fact, the hybridization gap of 
1.52 eV between (Ar+$\ful$)$3p$ and (Ar-$\ful$)$3p$ in our calculation is in good agreement with the measured value of 
1.6$\pm$0.2 eV \cite{morscher2010strong}. We use the symbol $n\ell$@ to denote the levels of the confined atom and 
@$n\ell$ to represent the levels of the doped $\ful$.
\begin{figure*}[ht]
\includegraphics[width=13cm]{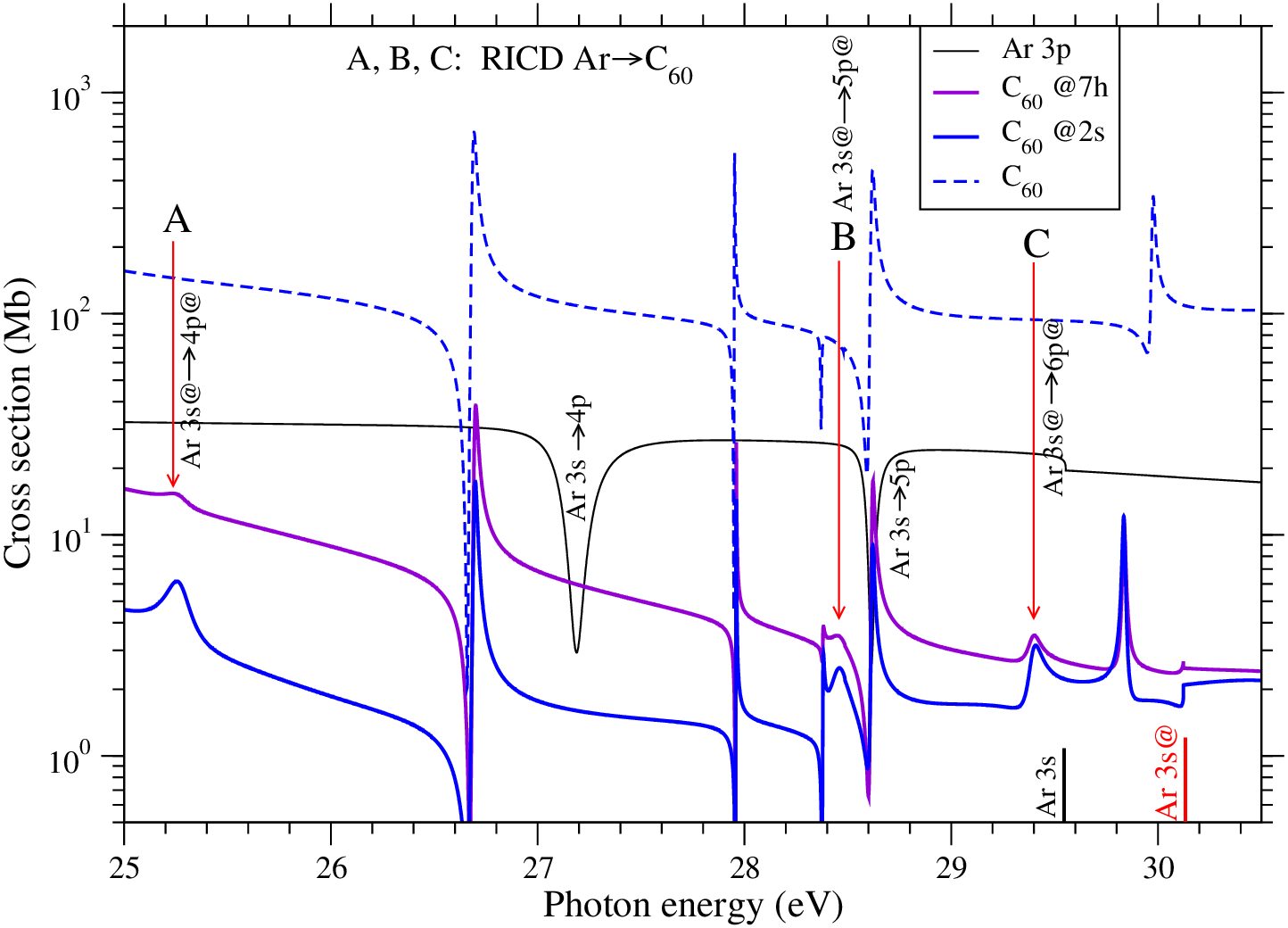}
	\caption{(Color online) Photoionization cross sections of free Ar $3p$ and empty $\ful$ compared with the results for $\ful$ @$7h$ and @$2s$
	levels in Ar@$\ful$. Three Ar-to-$\ful$ ICD resonances (labeled as A,B,C) amongst regular autoionizing resonances are identified in the $\ful$ @$7h$ and @$2s$ cross sections.}
	\label{figure2}
\end{figure*}

Figure 2 shows the 3p photoionization cross section of free Ar calculated using TDLDA. Two Auger window-resonances at 27.2 eV and 28.6 eV correspond to
regular autoionizing states formed by two lowest innershell excitations $3s\rightarrow4p,5p$. We also present in Fig.\,2 the cross sections for $\ful$ 
@$7h$, which is the highest occupied (HOMO) level of $\ful$ $\pi$ symmetry (one radial node), and
for @$2s$, which is the state at the bottom of the $\pi$ band. Both these cross sections exhibit a host of routine autoionizing
resonances corresponding to $\ful$ innershell excitations which also appear in the $\ful$ total cross section (shown)
at about the same energies. Three rather weak features, labeled as A, B, and C in Fig.\,2, are noted in the @$7h$ and @$2s$ 
curves which do not have partners in the free $\ful$ cross section, however. These are Ar-to-$\ful$ ICD resonances, resulting from the decay of
Ar $3s$@ vacancies from $3s@\rightarrow 3p@,4p@,5p@$ excitations, through $\ful$ @$7h$ and @$2s$ continua. Slight red-shifts 
in the position of these resonances compared to their free Ar 
counterparts are due to some adjustments in $3s$ ground and $np$ excited energies arising from confinement. Note that these
structures are from participant RICD processes only, as the spectator RICD is not included within TDLDA.

Figure 3 displays cross sections, over the same energy range of Fig.\,2, for the endofullerene hybrid levels, (Ar$\pm$$\ful$) $3p$. 
Features A, B, and C in these curves are resonances that emerge from the decay of $3s@\rightarrow 3p@,4p@,5p@$ excitations 
through the continuum of these hybrid levels. These features are similar in shape to the autoionizing resonances in free Ar $3p$ (included in Fig.\,3)
and appear at the same Ar-to-$\ful$ RICD energies (Fig.\,2). Remarkably, they are significantly stronger, particularly for (Ar-$\ful$) $3p$, 
than the Ar-to-$\ful$ RICDs. Another dramatic effect can be noted: The empty $\ful$ $3p$ cross section in Fig.\,3 shows autoionizing resonances 
corresponding to Auger decays of $\ful$ innershell vacancies. But the structures at the corresponding energies in hybrid channels from the decay of $\ful$ vacancies 
through the hybrid continuum are order of magnitude larger than the autoionizing resonances in empty $\ful$. We particularly identify the resonances labeled 
as 1 to 4 in Fig.\,3. In essence, Ar and $\ful$ innershell vacancies decay significantly more powerfully through the photoionization continua of
Ar-$\ful$ hybrid levels than they do through the continua of pure $\ful$ levels. These resonances are qualitatively different than the standard RICD. We 
show below that they emerge from a coherent interference between resonant Auger and intercoulombic channels that produce {\em divided} vacancies 
in the final state, vacancies shared by the confined atom and the confining fullerene. 
\begin{figure*}
\includegraphics[width=13cm]{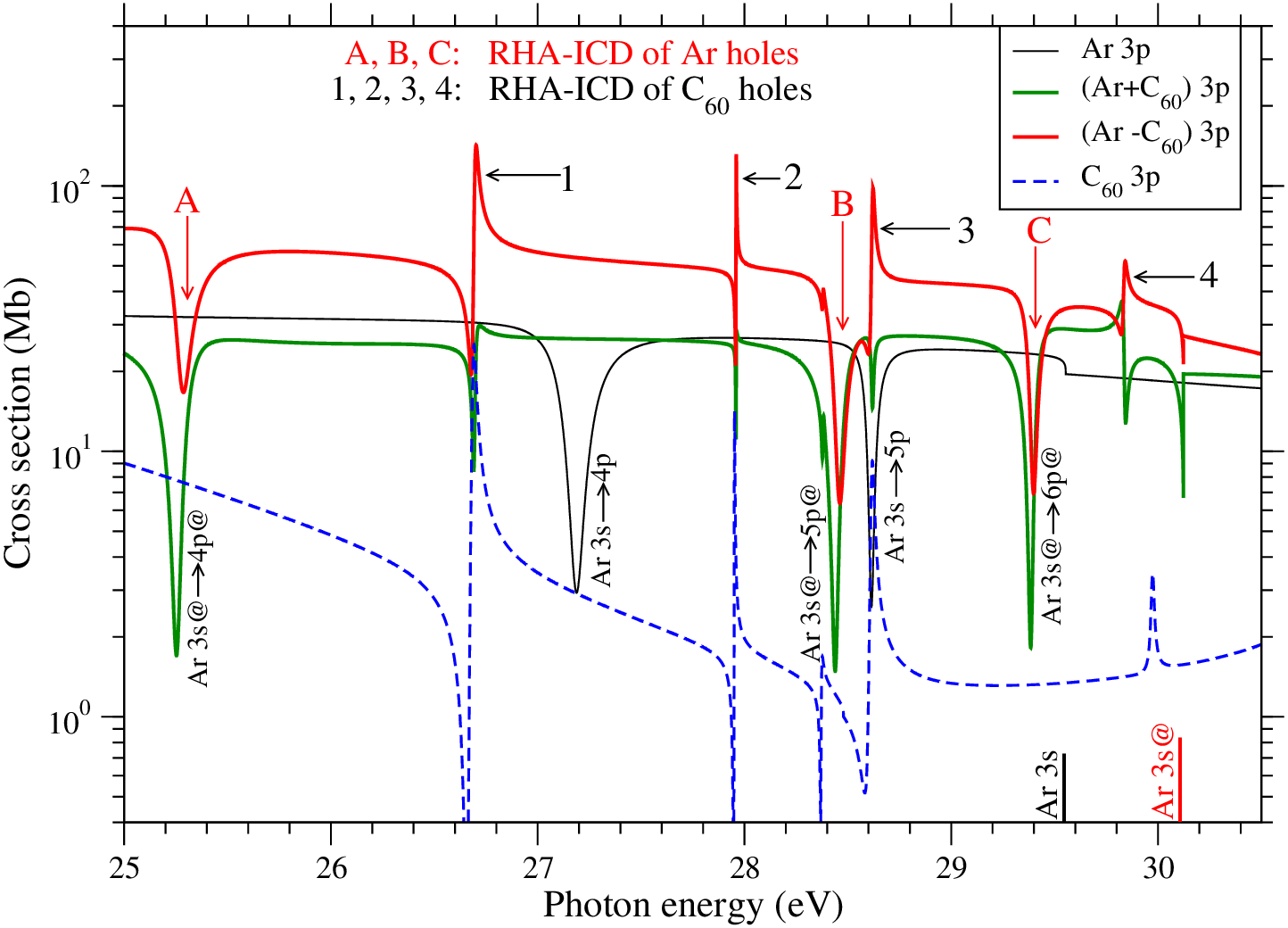}
	\caption{(Color online) Photoionization cross sections of free Ar $3p$ and $\ful$ $3p$ levels compared with those of their hybrid pair.}
	\label{figure3}
\end{figure*}

The TDLDA matrix elements for the dipole photoionization of (Ar$\pm$$\ful$) $3p$ levels, in the perturbative interchannel coupling framework introduced by Fano \cite{fano1961}, 
can be written as \cite{javani12alkaline-earth},
\begin{eqnarray}\label{gen-mat-element}
{\cal M}_\pm (E) &=& {\cal D}_\pm (E) + {M}^{c-c}_\pm (E)+{M}^{d-c}_\pm (E),
\end{eqnarray}
where the single electron (LDA) matrix element ${\cal D}_\pm (E) = \langle ks(d)|z|\phi_\pm\rangle$; ${M}^{c-c}$ and ${M}^{d-c}$ are respectively corrections
from continuum-continuum and bound-continuum channel couplings. ${M}^{c-c}$ constitutes the many-body contribution of relatively smooth nonresonant 
ionization cross section\cite{maser2012zn-hybrid}, while the resonance structures originate from ${M}^{d-c}$. Following Ref.\,[\onlinecite{fano1961}],
\begin{eqnarray}\label{dc-mat-element}
 {M}^{d-c} &=& \displaystyle\sum_{n\ell} \sum_{\eta\lambda} \frac{\langle\psi_{n\ell\rightarrow\eta\lambda}|\frac{1}{|{\bf r}_{\pm}-{\bf r}_{n\ell}|}
|\psi_{\pm}(E)\rangle}{E-E_{n\ell\rightarrow\eta\lambda}} {\cal D}_{n\ell\rightarrow\eta\lambda},
\end{eqnarray}
in which the $|\psi\rangle$ refer to interacting discrete $n\ell\rightarrow\eta\lambda$ and continuum (Ar$\pm$$\ful$) $3p\rightarrow ks(d)$ {\em channel} wavefunctions; 
$E_{n\ell\rightarrow\eta\lambda}$ and ${\cal D}_{n\ell\rightarrow\eta\lambda}$ are LDA bound-to-bound excitation energies and matrix elements, respectively. 
The excited states of the endofullerene are found to be hybridized, 
implying that innershell electrons from pure levels are excited to the hybrid levels. But we do not
expect significant differences in ${\cal D}_{3s\rightarrow\eta p}$ from this effect between free and confined Ar. This is because, even though hybrid excited
waves develop structures at $\ful$ shell, the Ar $3s$ wavefunction continues to localize on Ar (Fig.\,1), qualitatively unaffecting the overlaps.
Obviously, an identical reason also ensures practically unchanged $\ful$ inner excitation matrix elements from the doping.

Following Eqs.\,(\ref{bound-hyb}), the hybridization of the continuum channels in \eq{dc-mat-element} assumes the form
\begin{equation}\label{channel-hyb}
|\psi_\pm\rangle = \sqrt{\alpha}|\psi_{3p@ \scriptsize{\mbox{Ar}}}\rangle \pm \sqrt{1-\alpha}|\psi_{@3p \scriptsize{\mbox{C}_{60}}}\rangle.
\end{equation}
In Eqs.\,(\ref{channel-hyb}) we used @ to indicate the inclusion of the modifications of the continuum waves of the confined Ar and doped $\ful$.
Using Eq.\ (\ref{channel-hyb}) in \eq{dc-mat-element}, and recognizing that the overlap between a pure Ar bound state and a pure $\ful$ bound state 
is negligible, we can separate the atomic and fullerene regions of integration to obtain 
\begin{widetext}
\begin{eqnarray}\label{dc-mat-element2}
{M}^{d-c}_\pm (E) &=& \displaystyle\sum_{n\ell} \sum_{\eta\lambda}\left[\sqrt{\alpha}\frac{\langle\psi_{n\ell\rightarrow\eta\lambda}|\frac{1}{|{\bf r}_{\pm}-{\bf r}_{n\ell}|}
|\psi_{3p@ Ar}(E)\rangle}{E-E_{n\ell\rightarrow\eta\lambda}}
               \pm \sqrt{1-\alpha} \frac{\langle\psi_{n\ell\rightarrow\eta\lambda}|\frac{1}{|{\bf r}_{\pm}-{\bf r}_{n\ell}|}
|\psi_{@3p \scriptsize{\mbox{C}_{60}}}(E)\rangle}{E-E_{n\ell\rightarrow\eta\lambda}} \right]{\cal D}_{n\ell\rightarrow\eta\lambda}.
\end{eqnarray}
\end{widetext}
Obviously, if $n\ell\rightarrow\eta\lambda$ produces Ar innershell holes, resulting to resonances A, B, and C in Fig.\,3, then the first term on right-hand-side of 
Eq.\,(\ref{dc-mat-element2}) represents the ordinary {\em intracoulombic} Auger decay in Ar, while the second term denotes the Ar-to-$\ful$ RICD. Conversely,
for $\ful$ inner vacancies (resonances 1-4 in Fig.\,3), the first and second term, respectively, present reverse RICD ($\ful$-to-Ar) and $\ful$ Auger processes.
The decays are shown schematically in Fig.\,1.

For the ionization cross sections, which involve the modulus squared of the amplitude, two important mechanisms play out: First, Auger and intercoulombic
decay pathways in Eq.\,(\ref{dc-mat-element2}) combine coherently to induce resonances, allowing the creation of shared outershell vacancies. Therefore,
this decay pathway can be called resonant hybrid Auger-intercoulombic decay (RHA-ICD). Note
that both the terms in Eq.\,(\ref{dc-mat-element2}) are large, owing to substantial overlaps between innershell bound states and 
(Ar$\pm$$\ful$)$3p$ hybrid wavefunctions. This partly explains why the features identified in Fig.\,3 are stronger than corresponding autoionizing
and ICD resonances. Second, the resonances in the matrix element ${M}^{d-c}_\pm$ also interfere with the nonresonant part ${\cal D}_\pm + {M}^{c-c}_\pm$, 
[Eq.\,(\ref{gen-mat-element})] which is generally stronger for hybrid levels than pure $\ful$ levels \cite{maser2012zn-hybrid}. This interference, consequently, 
enhances RHA-ICD resonances compared to their Auger partners in pure $\ful$ channels, as seen for structures 1-4 in Fig.\,3. The results exhibit
completely different resonance shapes for Ar-to-$\ful$ RICDs (Fig.\,2) compared to corresponding RHA-ICDs (Fig.\,3), although their lifetimes increase only slightly. 
Noticeably, the lifetime (130 fs) of the Auger feature 1 decreases to about 40 fs for the respective RHA-ICDs (Fig.\,3), while there is a strong 
shape-alteration for the feature 4. Lastly, hybrid final-state
vacancies may have unique consequences for the spectator type RHA-ICD: the post-decay repulsive force will considerably increase compared to RICD, since a half vacancy 
will reside too close to a full vacancy either on Ar or on $\ful$, allowing stronger fragmentation forces.

RICD systems are visualized as natural antenna-receiver 
pairs at the molecular scale \cite{trinter13HeNe} where the antenna couples to the incoming photon and transfers energy to the receiver to perform. 
RHA-ICD processes, predicted here, can enhance the efficiency of the ultimate output by enabling the antenna to
also contribute to the emission resonantly with the receiver through a quantum coherence. The effect may have significant utilization in 
nanoscale antenna technology \cite{novotny2012NanoAntenna}.

In conclusion then, we used the TDLDA methodology to calculate a variety of single-electron resonances in the photoionization of Ar@$\ful$. 
Ar-to-$\ful$ ICD resonances
are calculated for the first time. A different class of resonances decaying into atom-fullerene hybrid final state vacancies has been found
which arises from the interference of the intracoulomb autoionizing channel with an intrinsically connected intercoulomb channel. These resonances
are significantly stronger than both regular ICD and Auger resonances, which make them amenable for experimental detection. They are likely to exist generally 
in the ionization continuum of, not only atomic endofullerenes, but of molecules, nanodimers, and fullerene onions that support hybridized electrons as well.

\begin{acknowledgments}
The research is supported by NSF and DOE, Basic Energy Sciences.
\end{acknowledgments}


\end{document}